\documentclass[aps, 10pt, twocolumn,preprintnumbers,prd, amsmath,amssymb,a4page,notitlepage]{revtex4-1} %, ,10pt
\usepackage{graphicx}% Include figure files latexsym,array,enumerate,letter,numbers,
\usepackage{bm}% bold math
\usepackage{amssymb}
\usepackage{epsfig}
\usepackage{color}
\usepackage{mathtools}
\usepackage{cases}
\usepackage{booktabs}
 \usepackage{url}

\usepackage[
colorlinks=true,
filecolor=black,
anchorcolor=blue,
linkcolor=blue,
citecolor=cyan, %red,
urlcolor=cyan,
linktocpage=true,
plainpages=false,
breaklinks=true,
pdfstartview=FitH
]{hyperref}

 \newcommand{\cs}[1]{\textcolor{black}{#1}}

\def\m{{\mu_0}}
\def\e{{\epsilon_0}}
\def\ga{g_a}

\newcommand{\be}{\begin{equation}}
\newcommand{\ee}{\end{equation}}
\newcommand{\bea}{\begin{eqnarray}}
\newcommand{\eea}{\end{eqnarray}}
\newcommand{\vev}[1]{\left<{#1}\right>}

\begin{document}

\title{Resonant Electric Probe to Axionic Dark Matter}

\author{Junxi Duan$^{1,2}$}
%\email{your@email, \& please check affliations}
\author{Yu Gao$^{3}$}
%\email{gaoyu@ihep.ac.cn}
\author{Chang-Yin Ji$^{1}$}
%\email{your@email,\& please check affliations}
\author{Sichun Sun$^{4}$}
%\email{sichunssun@bit.edu.cn}
\author{Yugui Yao$^{1,2}$}
%\email{your@email,\& please check affliations}
\author{Yun-Long Zhang$^{5,6,7}$}
%\email{zhangyunlong@nao.cas.cn}

\affiliation{
$^{1}$Centre for Quantum Physics, Key Laboratory of Advanced Optoelectronic Quantum Architecture and Measurement(MOE),
School of Physics, Beijing Institute of Technology, Beijing, 100081, China}
\affiliation{$^{2}$Beijing Key Lab of Nanophotonics \& Ultrafine Optoelectronic Systems,
School of Physics, Beijing Institute of Technology, Beijing, 100081, China}
\affiliation{$^{3}$Key Laboratory of Particle Astrophysics, Institute of High Energy Physics,
Chinese Academy of Sciences, Beijing 100049, China}
\affiliation{$^4$School of Physics, Beijing Institute of Technology, Beijing, 100081, China}
\affiliation{$^5$National Astronomy Observatories, Chinese Academy of Science, Beijing, 100101, China}
\affiliation{$^6$School of Fundamental Physics and Mathematical Sciences, Hangzhou Institute for Advanced Study, University of Chinese Academy of Sciences, Hangzhou 310024, China}
\affiliation{$^7$International Center for Theoretical Physics Asia-Pacific, Beijing/Hangzhou, China
}

\begin{abstract}
The oscillating light axion field is known as wave dark matter. We propose an LC-resonance enhanced detection of the narrow band electric signals induced by the axion dark matter using a solenoid magnet facility.
We provide full 3D electromagnetic simulation results for the signal electric field. %in Beijing Institute
The electric signal is enhanced by the high $Q$-factor of a resonant LC circuit and then amplified and detected by the state-of-the-art cryogenic electrical transport measurement technique. 
%The amplifier noise is the leading noise in the setup. We demonstrate that the setup has promising sensitivity for axionic dark matter with mass $m_a$ below $10^{-6}$ eV. 
\cs{The cryogenic amplifier noise is the dominant noise source in the proposed detection system. We estimate that the detection system can have a promising sensitivity to axion dark matter with mass below $10^{-6}$ eV.}
The projected sensitivities improve with the size of the magnetic field, and the electric signal measurement can be potentially sensitive to the quantum chromodynamics (QCD) axion with $g_{a\gamma} \sim 10^{-16}$ GeV$^{-1}$
\cs{ around $m_a  \sim 10^{-8}$eV,} with a multi-meter scale magnetized region.
\cs{This limit is around five orders of magnitude below the current constraint from the cosmic rays. }

%1. We investigate the axion dark matter sensitivity at BIT's electric measurement technique and magnetic solenoid facility.
%2. We provide EM simulation results for the electric signal inside the solenoid magnetic field.

%{\bf{Keywords}:}  Axion detection; dark matter; quantum transport; cryogenic technology.
\end{abstract}
\maketitle

\tableofcontents

{\section{Introduction}} 
\label{sect:intro}
The cold dark matter makes up around one-fourth of the Universe's total energy contents from cosmological and astronomical observations\cite{Ade:2015xua}. Except for the weakly interacting massive particles (WIMPs), one of the promising dark matter candidates is the well-motivated axion~\cite{Peccei:1977hh, Weinberg:1977ma, Wilczek:1977pj, Peccei:1977ur, Kim:1979if}. As a natural solution to the `Strong CP problem'~\cite{Crewther:1979pi} in quantum chromodynamics (QCD) ~\cite{Peccei:2006as,Kim:2008hd}, the axion emerges as the Nambu-Goldstone boson of a global $U(1)_{\rm PQ}$ symmetry extension of the Standard Model (SM), which is broken spontaneously at high scale. 
At the strong interaction's confinement scale the shift symmetry is broken by the QCD instanton potential~\cite{Callan:1977gz, Vafa:1984xg}, and the axion acquires a tiny mass and becomes a pseudo-Nambu-Goldstone particle. Cosmologically, the axion field becomes dynamical after the QCD phase transition, and the axion field's classical oscillation can account for the Universe's dark matter relic density~\cite{Turner:1983he}, a.k.a. the `misalignment' mechanism~\cite{Preskill:1982cy, Abbott:1982af, Dine:1982ah,Sikivie:1982qv,Ipser:1983mw,Sikivie:2006ni}. The QCD axion's typical parameter window is $m_a\sim {\cal O}(10^{-5}-10^{-3})$~eV~\cite{Sikivie:2006ni} or $m_a<{\cal O}(10^{-7})$~eV~\cite{Hertzberg:2008wr}. %Generalized axion-like particles 
In high-scale `invisible axion' scenarios such as the Dine-Fischler-Srednicki-Zhitnitsky (DFSZ)~\cite{Zhitnitsky:1980tq, Dine:1981rt} and the Kim-Shifman-Vainshtein-Zakharov (KSVZ)~\cite{Kim:1979if, Shifman:1979if} models,
QCD axions obtain the characteristic axion-gluon 
coupling $\frac{a}{f_a}\epsilon^{\mu\nu\rho\sigma}G_{\mu\nu}^c {G_{\rho\sigma}^c}$ via $SU(3)_c$-charged fermion loops, where $a$ is the axion field, $f_a$ is the axion decay constant, and $G_{\mu\nu}^c$ are the strengths of gluon fields.

General axion-like particles (ALPs) are light bosons that have the electromagnetic $a\vec E \cdot \vec B$ coupling yet may not necessarily couple to gluons. ALPs are abundant in string-theory motivated models~\cite{Svrcek:2006yi}; ALPs generally have relaxed $m_a-f_a$ relations, different from QCD axion and they can have a much wider mass range as the dark matter candidate. We will use `axion' to denote both the QCD axion and ALPs.% in the rest of this paper.
The axion may also couple to other gauge fields, in particular, the quantum electrodynamics (QED) coupling is crucial for low-scale searches, with the interaction Lagrangian term as
%\be
${\cal L}_{a\gamma\gamma}=-g_{a\gamma}a\vec E\cdot \vec B$,
%\ee
where $\vec{E}$ and $\vec{B}$ are the common electromagnetic field strengths. The  axion-photon coupling $g_{a\gamma}=c_{\gamma}\alpha_{\rm QED} / (\pi f_a)$ is dimension $-1$ and it is suppressed by  the $U(1)_{\rm PQ}$ scale $f_a$.
Here $\alpha_{\rm QED}=1/137$ is the fine-structure constant and the $c_{\gamma}$ coefficient depends on the particular UV model. 

The interaction in
${\cal L}_{a\gamma\gamma}$ motivates different search strategies that axions convert into photons inside the laboratory's electromagnetic fields~\cite{Primakoff:1951iae,Sikivie:1983ip}. People have been actively searching for axions by cavity haloscopes in the typical frequency window, including ADMX~\cite{ADMX:2021nhd}, HAYSTAC~\cite{Zhong:2018rsr}, CAPP~\cite{CAPP:2020utb},QUAX-$a\gamma$~\cite{Alesini:2020vny}, TASEH~\cite{taseh}, DANCE~\cite{Oshima:2021irp}, ORGAN~\cite{McAllister:2018xgn}, CAST-RADES~\cite{CAST:2020rlf}, non-cavity high-frequency designs like MADMAX~\cite{Li:2021mep,Knirck:2019eug}, BREAD~\cite{BREAD:2021tpx},TOORAD~\cite{Marsh:2018dlj,Schutte-Engel:2021bqm} etc. The cavity haloscope design achieved the highest sensitivity so far due to the resonance cavity's high-$Q$ enhancement~\cite{Sikivie:1983ip,Yang:2022uil} that matches the axion dark matter's energy dispersion. 
At lower axion masses with an increased signal photon wavelength, non-cavity measurements become a viable option, such as cryogenic searches for conversion in a strong magnetic field, e.g., ADMX-SLIC~\cite{ADMXSLIC}, DM-Radio~\cite{Brouwer:2022bwo}, BASE~\cite{BASE}, ABRACADABRA~\cite{Salemi:2021gck,Kahn:2016aff}, as well as conversions in a strong electric field~\cite{Gao:2020sjn,Gao:2022zxc}, where alternative narrow-band filtering, e.g. an electronic LC circuit~\cite{Sikivie:2013laa}, can replace the cavity to provide front-end high $Q$-factor enhancement. See Ref.~\cite{Irastorza:2018dyq,Sikivie:2020zpn} for recent reviews of different axion search strategies. 

%In this {\it Letter}, 
Here we investigate the prospective narrow-band $g_{a\gamma}$ sensitivity of an LC-resonant measurement of axion dark matter-induced electric signals inside an ${\cal O}(10)$ Tesla magnetic field. Borrowing from the modern quantum transport experimental techniques, narrow band measurement can be done by cryogenic amplifiers and phase-lock readout. \cs{Our proposed detection system} is straightforward but stands as one of the few cases to propose an electric resonant phase-sensitive measurement to achieve great sensitivities, compared to the popular measurements of tiny magnetic signals inside the strong external magnetic background field e.g. in ABRACADABRA, ADMX-SLIC, DM-Radio, etc. We name our experimental proposal as {\bf{Resonant ELEctric Axion Probe (ReLEAP)}}. Previously, broad-band axion-induced voltage signal has been studied inside a solenoid~\cite{McAllister:2018ndu}, and also in a magnetic torus~\cite{Tobar:2020kmz}. The field distributions induced by effective currents for long solenoid geometries are computed in recent theory studies~\cite{Beutter:2018xfx, Ouellet:2018nfr}. We perform full numerical electromagnetic (EM) simulations to calculate the solutions of axion-induced electric signals in finite solenoid dimensions. We show that high $g_{a\gamma}$ sensitivity can be achieved by high $Q$-factor enhancement for a multi-centimeter magnetized volume. Probing the QCD axion parameter space requires a multi-meter scale magnetized region with state-of-art techniques.

\bigskip
\bigskip
%\vspace{10pt}
{\section{Long Solenoid Solutions}} 
\label{sect:em}
The dark matter axion field near the Earth is modeled by 
\be\label{axion1}
a(x,t)\approx a_0{\rm cos}\left(m_a\vec v\cdot \vec x-\omega t +\phi_0\right),
\ee
where $\omega= m_a\left(1+{1\over 2}v^2\right)$ and $v\sim 10^{-3}c$ is the solar system's relative velocity to the Galactic dark matter halo. The magnitude $a_0={\sqrt{2\rho_{\rm DM}}/ m_a}$ and $\rho_{\rm DM}=0.4$ GeV cm$^{-3}$ is the local dark matter density \cite{Read:2014qva}.
\cs{
Due to the non-relativistic velocity, low mass axion dark matter can have a long de-Broglie wavelength $\lambda_{a}=2\pi/(m_a v_a)$ that exceeds laboratory dimensions. 
The axion signal coherence time scale $\tau_a=\lambda_a/\delta v_a\sim 2\pi/(m_av_a^2)$ is much longer than the period of the converted photon $\tau_\gamma=m_a^{-1}$, 
and $v_a$ is the velocity of axion like particles.
Thus, the dark matter axion field is often regarded as a non-relativistic, monochromatic plane wave with a small frequency spread  $Q_a^{-1}=\delta f/f\sim v_a^{2}$. 
In typical dark matter halo models $\delta v_a\sim v_a$ and $Q_a\sim 10^{6}$. } 
%And higher level of coherence is possible in halo models with $\delta v_a\ll v_a$~\cite{Sikivie:2001fg, Armendariz-Picon:2013jej}.
%Under our laboratory conditions, the axion dark matter field is regarded as an external background field with an oscillation frequency $f=2\pi m_a$ and a wave-vector $\vec{k}_a = m_a \vec{v}_a$, of which direction in the lab frame depends on the local relative velocity to the dark matter halo and the experimental apparatus orientation. 
This field enters the EM equation as a small perturbation source.

The axion-modified Maxwell's equations read as
\be
\begin{array}{ll}\qquad\quad
\vec\nabla\cdot \vec E
=\rho_e-g_{a\gamma}\vec B\cdot \vec\nabla a , \\
\vec\nabla\times\vec B-{\partial \vec E \over \partial t}
=\vec{j}_e -g_{a\gamma}\left(\vec E\times \vec \nabla a- \vec B{\partial a\over  \partial t}\right) ,
\end{array}
\ee
which observe additional sources from photon coupling to the background axion field. Here $\{\rho_e, \vec{j}_e\}$ denote the SM electric current in conventional electromagnetism. Under a static magnetic field, these new sources include the velocity-dependent effective charge density $\rho_a =-g_{a\gamma}\vec B\cdot \vec\nabla a$, and the effective displacement current density $\vec{j}_a= g_{a\gamma} \vec{B}{\partial a\over  \partial t}$. Both $\rho_a$ and $\vec{j}_a$ oscillate with the axion field in \eqref{axion1}. 
To the leading order of $g_{a\gamma}$, the set of EM equations can be simplified,  
\iffalse
into
\bea
\vec\nabla\cdot \vec E&=&
\rho_e+ \rho_a  \nonumber\\
\vec\nabla\times\vec B-{\partial \vec E \over \partial t}&=& \vec{j}_e + \vec{j}_a \nonumber\\
\vec\nabla\cdot\vec B&=&0\nonumber\\
\vec\nabla\times \vec E&=&-{\partial \vec B\over \partial t}~,
\eea
\fi
$\rho_a$ and $\vec{j}_a$ only take into account external EM fields other than those induced by axion oscillations.

 \cs{
We consider a 14 Tesla magnetic field with $\vec{B}=B_0 \hat{z}$ inside the solenoid of radius $R$. In the simplified long-solenoid geometry, the new source terms can be written as
\be
\begin{array}{ll}
 &\rho_a   =  g_{a\gamma}B_0 (\vec v\cdot \hat{z})\sqrt{2\rho_{\rm DM}}{\rm cos}\left(\omega t\right),\\
&\vec j_a= \hat{z} g_{a\gamma} B_0 (1+{1\over 2}v^2)\sqrt{2\rho_{\rm DM}}{\rm cos}\left(\omega t\right).
\end{array}
\ee
}
%where $\omega= m_a\left(1+{1\over 2}v^2\right)$.
%Their amplitudes are determined by the local dark matter density and are independent of $m_a$. Comparing with $\vec{j}_a$, the charge $\rho_a$ is ${\cal O}(10^{-3})$ suppressed by the solar system's non-relativistic velocity. 
The Earth's rotation modulates $\vec{v}$ direction by 24 hours unless our local relative velocity to the dark matter halo aligns with the Earth's spinning axis (geometric North). %At low frequency, when the experimental area is much smaller than the axion field wavelength $\lambda_a$, the laboratory $\vec{B}$ field only samples part of the axion field's period, and a $v^{-1}_a m_a$ modulation arises due to the non-relativistic relative motion of the axion field.
In principle, these slow modulations can be accounted for during data analysis, and here we consider a constant amplitude in simplified analytic calculations.
%The effective charge $\rho_a$ and current $\vec{j}_a$ are spatially modulated by the axion field wavelength along $\hat{k}_a$, and propagate along  $\hat{k}_a$ at speed $v_a$.  
$\rho_a$ and $\vec{j}_a$ satisfy local charge conservation condition
$
\partial_t \rho_a + \nabla \cdot \vec{j}_a =0,
$
yet their magnitudes both oscillate globally with $a(t)$. 
%Unlike ordinary matter with equal number of opposite-sign force carriers, $\rho_a$ carries a sign and the net charge does not average to zero on scales smaller than the axion field wavelength $\lambda_a$. 
%Wave-like behavior of $\rho_a,\vec{j}_a$ also leads to spatial modulation in the induced $\vec{E}_a$ field's $\hat{z}$ and radial components over $\lambda_a$.

Working in the long-wavelength limit ($\lambda =2\pi c/\omega \gg R$), we can consider a short ($l\ll \lambda $) cylindrical section of the solenoid, where the axion-induced electric field $\vec{E}_a$ can be considered as uniform along $\hat{z}$, and its relative difference at two ends of the section is less than ${\cal O}(4l/\lambda)$. 
%$\rho_a$ is nonzero and uniformly distributed in this small section ($r<R$), and $\rho_a=0$ outside the solenoid ($r>R$), under cylindrical coordinates.
 For the radial component $E_{a,r}$ of the induced electric field, we have
 \be 
{E}_{a,r} = 
-\cos{\omega t}~%\hat{r}  
\cdot
\left\{ 
\begin{array}{ll}
 {1\over 2} g_{a\gamma}B_0 v_z\sqrt{2\rho_{\rm DM}}r,&{r\le R}, \\
 {1\over 2}g_{a\gamma}B_0 v_z\sqrt{2\rho_{\rm DM}}\frac{R^2}{r},&{r\ge R}.
\end{array}
\right.
\ee
%as $E_{a,z}$ is in the same direction at the up and lower surfaces of the section and their contributions to the surface integral largely cancel.
\begin{figure}[h]
\includegraphics[scale=0.18]{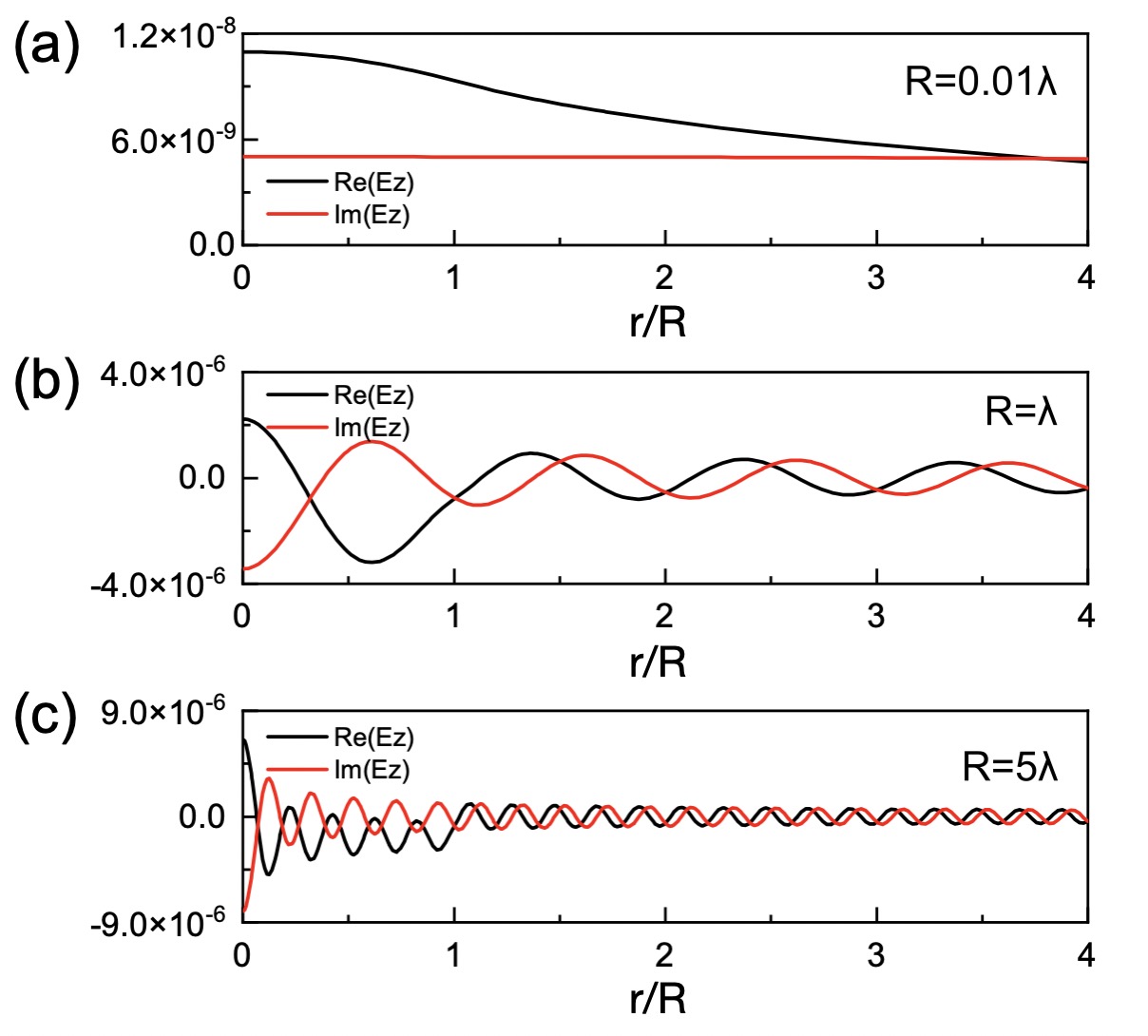}
\caption{Validation 2D plots by the electromagnetic simulation to compare with analytic ansatz. $E_z$ is in the unit of $V/m$, $R$ is the radius of the solenoid magnetic field and $\lambda=2\pi/m_a$.
}
\label{fig:sims2D}
\end{figure}

For the $\hat{z}$ component, an ansatz is given in Ref.~\cite{Ouellet:2018nfr} for $E_{a,z}$ induced by uniformly distributed $\vec{j}_a$ in the low-velocity limit (i.e. setting $\vec{\nabla} a \rightarrow 0$ that drops out the $\rho_a$ term), which reads 
\be 
E_{a,z} =  g_{a\gamma} a B_0 
e^{i\omega t} \cdot\left\{ 
\begin{array}{ll}
\alpha(\omega) J_0(\omega r) - 1,&~{r\le R}, \\
\beta(\omega) H^+_0(\omega r),&~{r\ge R}. 
\end{array}
\right.
\label{eq:Ez}
\ee
%where $ j_a/\omega = g_{a\gamma} a B_0$. 
Here $J$ is the first kind of Bessel function and $H$ denotes the Hankel function. The coefficients are given by
$\alpha(\omega) =  \frac{i \pi \omega R}{2}H^+_1(\omega R)$,
and $\beta(\omega)  = \frac{i \pi \omega R}{2}J_1(\omega R)$.
%\be \begin{array}{ll} \alpha(\omega) &=  \frac{i \pi \omega R}{2}H^+_1(\omega R),  \\
%\beta(\omega) &= \frac{i \pi \omega R}{2}J_1(\omega R). \end{array}\ee
%and note we factored out $j_a/\omega$ for later convenience. 
In Fig. \ref{fig:sims2D}, we perform an EM simulation as a validation test for the $E_z$ solution.

In the long-wavelength limit, or $\omega R\ll 1$, the magnitude of this $E_{a,z}$ solution is found to be geometrically limited by $(\omega R)^{2}$~\cite{Beutter:2018xfx, Ouellet:2018nfr}, 
%A few comments are due for the axion-induced electric field signals. 
%Since $\vec{E}_a$ is affected by $\partial_t \vec{j}_a$, it is unsurprising that $E_z$ is $r$-dependent. This can be seen from the nonzero integral $\oint \vec E\cdot {\rm d}\vec{l}$ along any vertical rectangular closed-path with its two vertical sides at $r=0$ and $r\neq 0$. While the contribution from the horizontal sides cancel, the rectangle encloses a time-variant magnetic flux induced by the $\partial_t \vec{j}_a$ distribution, so that $E_z(r)\neq E_z(0)$. 
which indicates that $E_z$ measurement is most sensitive to axion mass that matches the magnetized volume size, i.e. $m_a\sim {\cal O}(R^{-1})$. $E_r$ is $v_a$-suppressed and its measurement will require cylinder-shaped plates. For a given experimental size, the best $g_{a\gamma}$ sensitivity of  $E_z$ measurement is obtained at maximally allowed $m_a$.  Here we will consider $E_z$ measurement with horizontally placed parallel plates. Insertion of conductor plates will strongly modify the distribution of $\vec{E}$ components parallel to its surface, and less so for the perpendicular $\vec{E}$ components away from conductor edges.

\begin{figure}[ht]
\includegraphics[scale=0.16]{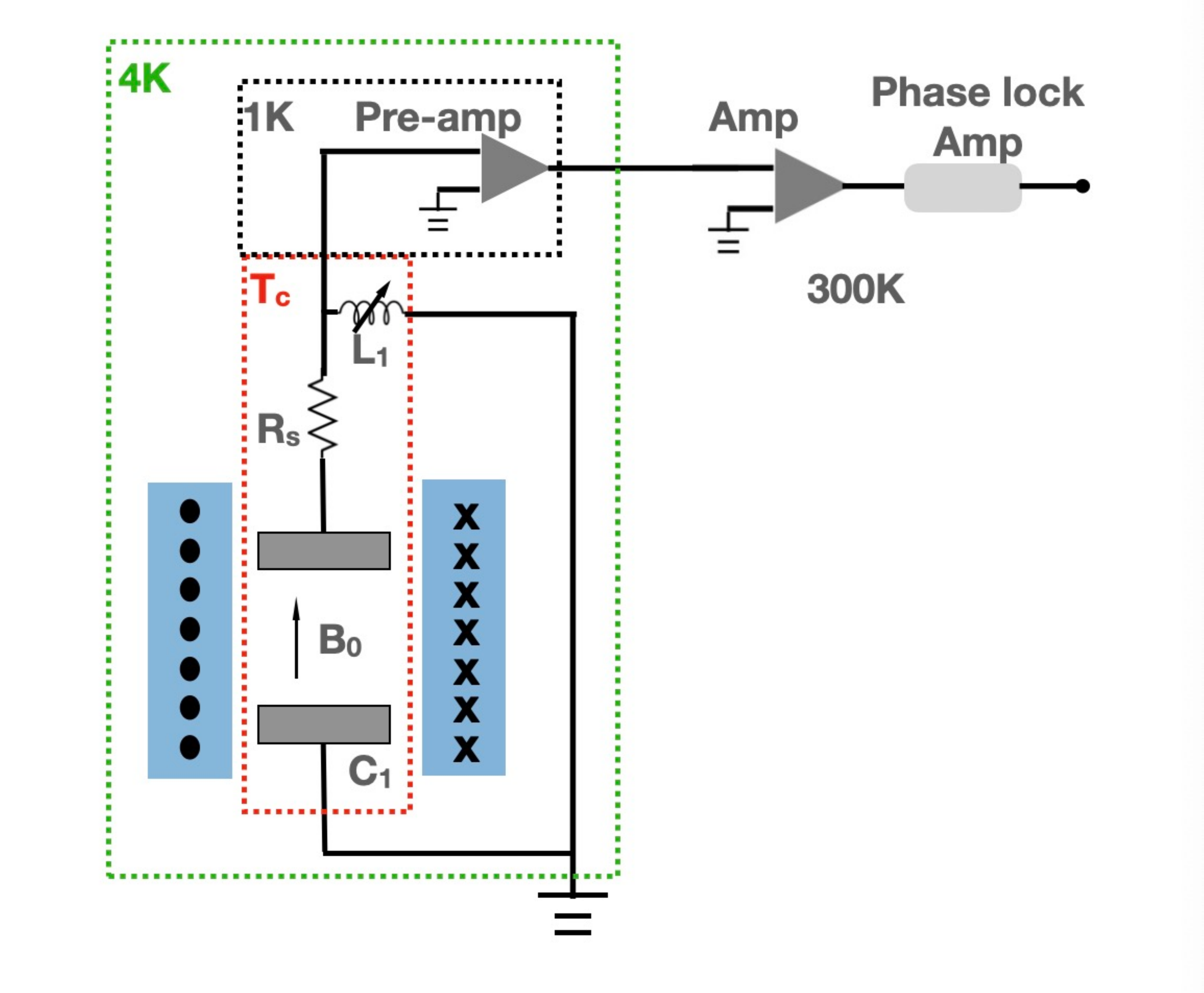}
\caption{Illustration for the experimental scheme. The voltage signal from parallel plates inside a solenoid magnetic field is enhanced by a resonant LCR circuit and then fed to amplifiers. The LCR is under $T_c \ll 1$ K to reduce thermal noise.}
\label{fig:setup}
\end{figure}

\bigskip
%\vspace{10pt}
{\section{Proposed Experimental Setup}} 
\label{sect:setup}
The proposed experimental setup is illustrated in Fig.~\ref{fig:setup}. Integrating the electric field around the plate's surface gives the charge accumulation $ q=\int \vec{E}\cdot {\rm d}\vec{A}$, and the field distribution $\vec{E}(x,t)$ is obtained by EM simulation. As $q(t)$ oscillates with the DM axion field, the signal current in the circuit connected to the plates is $I_a(t)=\dot{q}$. 
%We refer to the Appendix for theoretical backgrounds and simulation details.

%The typical experimental setup is illustrated in Fig.~\ref{fig:setup}. Integrating the electric field around the plate's surface gives the charge accumulation
%\be 
%q=\int \vec{E}\cdot {\rm d}\vec{A}
%\ee
%and $\vec{E}$ distribution is obtained via EM simulation. $q(t)$ oscillates with the DM axion field and $q_0$ denotes its magnitude. The signal current in the circuit connected to the plates is then $I_a(t)=\dot{q}$ with its magnitude equals $I_0^{\rm max}=\omega q_0$.

The parallel plates are connected in an LCR circuit tuned to its resonance point to provide enhancement~\cite{Sikivie:2013laa} to the signal. The effective capacitance of the parallel plates was obtained by comparing $q$ with the plates' voltage difference. %Here we would assume the plates' capacitance is larger than wiring's stray capacitance, and denote the effective total capacitance in the circuit as $C$. 
The LCR circuit will contain adjustable inductor(s) $L_1$ in order to: (i) ensure resonance at the axion field frequency $f=\omega/(2\pi)$; (ii) compensate for stray inductance from wiring; (iii) meet impedance matching with amplifiers and read-out. The total induction in the LCR loop is $L = L_1 + L_D$,
%\be ,\ee
where $L_D$ denotes any additional inductance from coupling to the amplifiers and readout. LCR resonance requires at $\omega^{-2}=LC$ with the quality factor in signal power as $Q_c$ that matches the axion field's $Q_a$, which is due to dark matter's kinetic energy dispersion. Usually, \cs{$Q_a = 1/\delta(v^2)\sim 10^{6}$} in conventional virialized dark matter distributions, while sharper values $Q_a$ are also possible~\cite{Armendariz-Picon:2013jej,Sikivie:2001fg}. Quality factor matching requires $Q_c \sim Q_a$, such as $Q_a/Q_c=1$ for max signal power, and typically $Q_c/Q_a=O(1)$ to balance between signal power and overall scan rate. More precisely, the matched $Q_c$ should be replaced by the system's `loaded’ quality factor~\cite{Brubaker:2017ohw} after being coupled to the amplifier/detector chain, yet here it suffices to use $Q_c\sim Q_a\sim 10^6$ as an estimate.

Working on the LCR resonance point, the saturated signal current driven by the axion field is
\be 
I_{a} =  Q_{c}\cdot q_0 \omega \cos(\omega t),
\ee
where $q_0$ is the electric charge magnitude. The signal dissipation power in the circuit is $I_a^2 R_s$ and it is capped by the DM signal's energy input,  e.g. the axion-photon conversion rate. 

\bigskip

{\subsection{The geometric factor and numerical simulation}} 

The geometric factor $\eta(\omega)$ is defined as
\be 
\eta(\omega) \equiv \frac{\int \vec{E}_a\cdot {\rm d}\vec{A}}{\int g_{a\gamma}a\vec{B}_0\cdot {\rm d}\vec{A}}
\label{eq:eta}
\ee
with the plate's surface charge buildup due to the signal $\vec{E}_{a}$ normalized by the denominator integral, whose integrand $g_{a\gamma} a B_0$ represents the theoretical amount of axion-induced field-strength mixing and is conveniently expressed in terms of the experiment's $B$-field strength and the axion-photon coupling. $\vec{E}_a$ follows the long-solenoid ansatz, and it is $\omega$ dependent. 
With the long-solenoid solution of $E_z(r)$ from Eq.~\eqref{eq:Ez}, the geometric factor evaluates to
\bea 
\eta(\omega) &\approx&  \frac{1}{\pi R^2}\left|\int_0^R [\alpha(\omega)J_0(\omega r)-1] \cdot 2\pi r{\rm d}r\right| \nonumber \\
 & =& \left| i\pi J_1(\omega R)H_1^+(\omega R) - 1 \right| 
\eea
by ignoring field distortion on plate edges, where 
%$x_0\equiv \omega R$ and 
$J_{0}$ and $J_1$ are first kind Bessel functions,
$H_1^+$ denotes the Hankel function.
This quantity is of the order $\eta\approx 0.1$ near the cutoff $\omega R = 1/4$. 
For signal coherence, we require the converted photon wavelength to be longer than the apparatus size, so that
$\lambda  = 2\pi m_a^{-1} \gg 2\cdot {\rm max}(2R, d)$,  
where $d$ is the relevant distance along $\hat{z}$.
\iffalse\fi
%
Accurate $\eta(\omega)$ values including edge distortions can be obtained by EM simulation.

With a quality factor $Q_c$ and the resonance resistance $R_s=(Q_c\omega C)^{-1}$, the average dissipated `signal power' $P_{\rm sig}=\vev{I_a^2 R_s}$ can be written as
%$\vev{I_a^2 R_s} = \frac{(Q_c \omega q_0)^2}{2Q_c \omega C}=Q_c\cdot \left[ g_{a\gamma}^2\eta(\omega)^2 \cdot \rho_{\rm DM} B_0^2\right] \cdot \pi R^2 d/2$,
\bea  
P_{\rm sig}&=& \frac{(Q_c \omega q_0)^2}{2Q_c \omega C} \nonumber \\ & =& Q_c\cdot \left( g_{a\gamma}^2\eta(\omega)^2 \cdot \frac{\rho_{\rm DM}}{m_a} B_0^2\right) \cdot \pi R^2 d, 
\label{eq:sig_power}
\eea
in which $d$ is the distance between plates, and we have used $C\sim \pi R^2/d$ as the low-frequency estimate for the plates' capacitance. % (see Appendix for justification). 
Note this formula resembles the signal power formula for a cavity haloscope in the sense that $\pi R^2 d$ represents the enclosed magnetized volume between the plates, where axion converts to photon coherently.  We use a geometric form factor $\eta$ to relate the plate's charge buildup $q_0$ and the applied magnetic field $B_0$.

\begin{figure}[ht]
\includegraphics[scale=0.3]{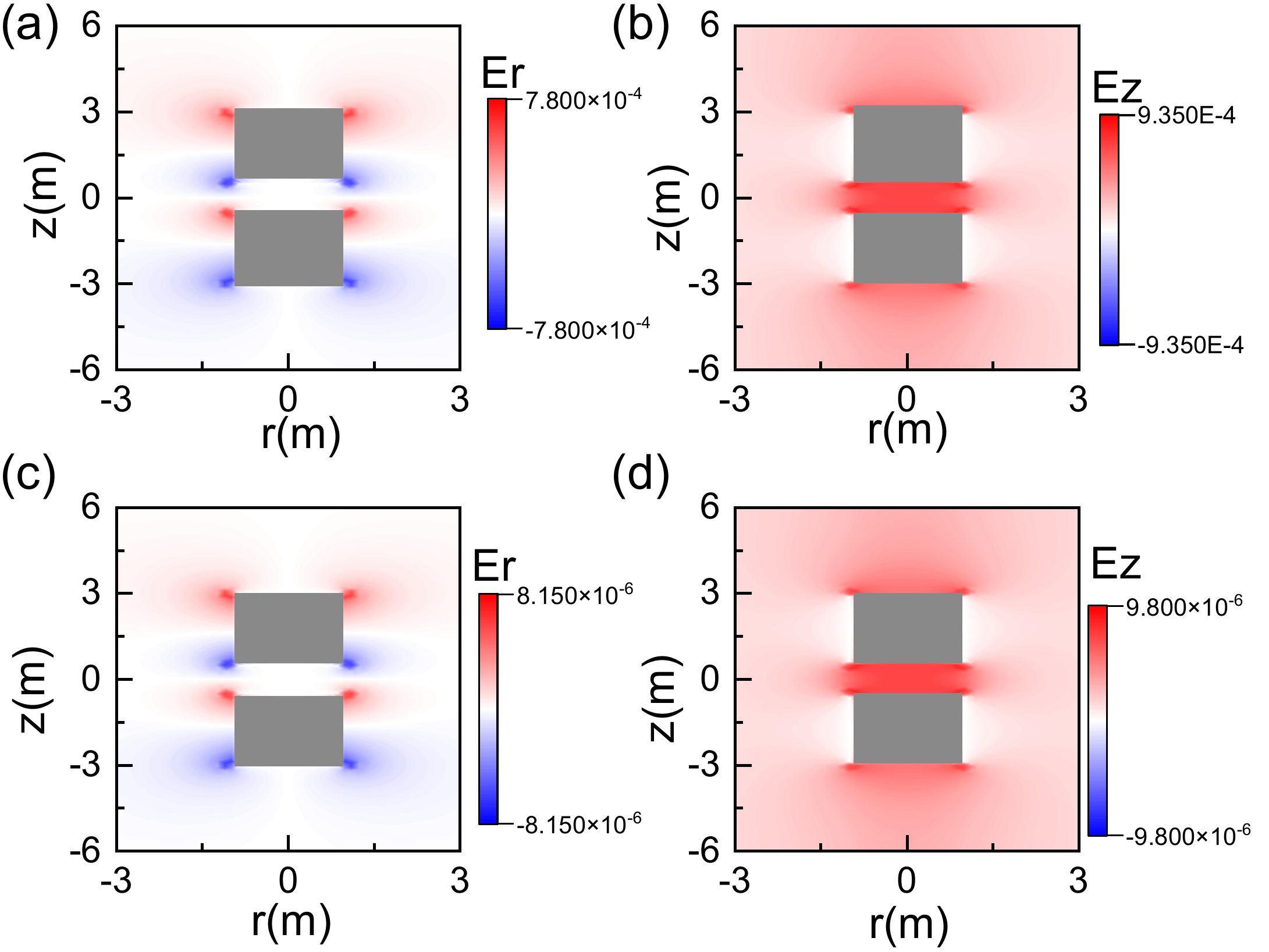}
\caption{3D simulations in a magnetic solenoid with conducting plates. The detector size is $R=1m, d=1m$. 
$E_z$ and $E_r$ are in the unit of $V/m$. The thick conducting plates are chosen to show well-separated upper and bottom surfaces. The top two plots are for $m_a= 10^{-8}$eV and the bottom two are for $m_a= 10^{-10}$eV.  %More plots can be found in appendix.
}
\label{fig:sims3D}
\end{figure}

Here we perform numerical electromagnetic simulation for the electric field signal $E_a$ around a solenoid. We introduce $\rho_a,\vec j_a$ as small external charge/current sources into the standard Maxwell equations and use {COMSOL} Multiphysics package \cite{COMSOL} as the EM simulator. 
We consider a simplified 2D long-solenoid simulation as validation for the analytic solutions in Fig.~\ref{fig:sims2D}, and then a finite-size 3D simulation for a $14$ Tesla solenoid of different sizes in Fig.~\ref{fig:sims3D}, with metallic plates inserted for electric charge build-up and signal readout. 
%Simulation setup details and 2D results are discussed in Appendix.
%\ckk{\bf Figures for $E^{a}_r$ and $E_z^{a}$, with the choice of benchmark frequencies.}
We can calculate the accumulated charge on the plates numerically and the results are tabled in  Table \ref{tab:charge_on_plate} with different detector sizes. % The induced current $I\sim Q/f$ can be amplified by an LC circuit pickup with a matched resonance frequency, and further enhanced by amplifiers during signal readout. 
We include the effect of velocity and finite-size distortions on the edges, different from~\cite{Ouellet:2018nfr}.
We also refer to the conversion in \cite{Tobar:2018arx}, for the modified Maxwell's equations in the standard international(SI) unit.
%We also consider different axion wavelengths,  which yield different accumulated charges on the plates.

\begin{table}[h]
\begin{tabular}{c|c|c}
\hline
~$\omega R$~ & ~$2R:d = 2$m$:1$m~&  ~$2R:d = 2$cm$:5$cm~\\
\hline
1 & 1.81$ \times 10^{-2}$&1.48$ \times 10^{-2}$ \\
0.1 &1.87$ \times 10^{-4}$ &2.31$ \times 10^{-4}$\\
0.01 &5.04 $ \times 10^{-6}$&2.58$ \times 10^{-6}$ \\
\hline
\end{tabular}
\caption{EM Simulated value of the form factor $\eta(\omega)$ in Eq.~\eqref{eq:eta}, \cs{for $R$=1m, $d$=1m and $R$=1cm, $d$=5cm}, respectively. The frequencies are chosen according to $ \omega= \frac{2\pi}{0.124m}(m_a/10^{-5}$eV).}
\label{tab:charge_on_plate}
\end{table}

The magnetized region, parallel plates, and the \cs{LCR(inductor-capacitor-resistor)} circuit are placed under cryogenic temperature $T_c$. 
Typically the setup can be installed in a dilution refrigerator equipped with a superconducting magnet of field strength $\vec{B}_0$ up to 16 T. Environmental fluctuations like the Earth's magnetic field can be removed by proper shielding. The solenoid magnetic field's own instability in the measured bandwidth may be of concern and must be accounted for via in-situ measurements.

Two strategies are possible in signal measurement. (1) Inductively measure the signal current in the LCR circuit with a SQUID \cs{(superconducting quantum interference device).} With a flux transformer coupled to the LCR loop, the signal current generates an oscillating magnetic field that can be picked up with a SQUID. To minimize intervention from the solenoid, the SQUID can be protected by a lead box on the 4 K plate of the dilution refrigerator, where the magnetic field is less than 500 Oe and is below the critical field of lead. (2) Directly amplify the voltage signal in the LCR circuit by using a cryogenic amplifier, followed by phase-sensitive detection with narrow-band lock-in amplifiers. The latter method is shown in Fig.~\ref{fig:setup}, and it is a viable option in case $m_a$ is higher than the operating frequency range of available SQUIDs. In both methods, the signal feeds into a cryogenic amplifier and its added noise will affect the detection sensitivity.
%\bigskip

\bigskip
%\vspace{10pt}
{\subsection{Expected sensitivity}} 
\label{sect:sensitivity}
Major noise sources that affect the signal-to-noise ratio (SNR) include the thermal fluctuation in the LCR circuit and the noises introduced in the amplifier-detector chain. One usual noise bottleneck is the amplifier/detector noise, especially the input noise at the first cryogenic amplifier. 
\cs{Considering HEMT amplifiers, this amplifier noise is $P_{\rm AMP}=4k_B T_{\rm AMP} \Delta f\sim 4k_B T_{\rm AMP}\cdot\omega/(2\pi Q)$, where the detector bandwidth matches with the experiment's resonance quality factor. $T_{\rm AMP}$ is the noise's effective temperature and it is usually higher than the thermal white noise at the working temperature~\cite{4359086}. For instance, a typical voltage noise at $0.3$ nV/$\sqrt{\rm Hz}$ corresponds to $T_{\rm AMP}\sim 40$ K.} 
As both the signal and the LCR curcuit's thermal noise are amplified, the effective noise temperature is determined by the {\it greater} between the amplifier's input noise $P_{\rm AMP}$ and the amplified LCR thermal noise ${\rm G}\cdot 4 k_B T_c$:
\bea 
P_{\rm noise}/\Delta f  &\approx& {\rm G}\cdot 4 k_B T_c  +  4 k_B T_{\rm AMP}, \\
 & \equiv & {\rm G}\cdot k_B T_N \nonumber
\eea
%\ckk{[Duan: HEMT reference?]} 
where $T_c$ is the environmental temperature at the LCR circuit and the parallel plates, and G is the amplifier's gain. It is clear that a high amplifier gain would help overcome the noise contamination from $P_{\rm AMP}$. 
\cs{For a  cryogenic amplifier with ${\rm G}=30$ dB and  $T_{\rm AMP}= 40$ K~\cite{AMP,4359086}, the two noises cross-over is at $T_N=0.16$K and correspondingly $T_c=0.04$K. Thus we choose $T_N\sim 0.1$K as a reasonable experimental condition. 
As $Q_c\sim 10^6$, the detector bandwidth is also best narrowed down to around $\Delta f = Q_c^{-1} f$ to match that of the LCR circuit. With our $m_a$ mass range of interest a $\Delta f$ in the $10^{-3}$ Hz -$10$ Hz range can be readily achieved via phase-lock measurement~\cite{PSD}. }

\cs{For $T_N\ge0.1$K and $f< 0.1$GHz, quantum noise ${\cal O}(1)hf/k_B$ is below mK level and is subdominant to thermal noises. The electronic $1/f$ noise is small till at a very low frequency. 
Experimental systematics will include mechanic vibrations and magnet instability, etc., within the measured frequency bandwidth, as well as detector systematics like voltage and ground drifting~\cite{morris2001measurement}, etc. These systematics should be in-situ calibrated and we will only show a statistics-based sensitivity in this work.}

The signal-to-noise ratio ${\rm SNR}$
can be computed with Eq.~\eqref{eq:sig_power} as,
%$\frac{Q_c g_{a\gamma}^2 \eta^2 f_c^{-1}  \rho_{\rm DM} B_0^2 (\pi R^2 d)}{k_B T_{\rm N}}\sqrt{Q_c \frac{2\pi}{m_a^3}\Delta{t}}$,
\bea {\rm SNR} &=& \frac{P_{\rm sig.}}{k_{\rm B}T_{N}}\sqrt{\frac{\Delta{t}}{\Delta f}}  \\&=& \frac{Q_c g_{a\gamma}^2 \eta^2 f_c^{-1}  \rho_{\rm DM} B_0^2 (\pi R^2 d)}{k_B T_{\rm N}}\sqrt{Q_c \frac{2\pi}{m_a^3}\Delta{t}}.\nonumber\eea
where $\Delta t$ is the observation time,  $f_c\sim {\cal O}(1)$ is a coefficient for the signal phase-lag of the plates' capacitance, see Appendix for the capacitance on plates, and the amplifier gain is suppressed. Note the form factors $\eta$ and $f_c$ also depend on $m_a$.
We require SNR$=3$ as the sensitivity criterion and place a constraint on $g_{a\gamma}$,
\bea 
g_{a\gamma}^{\rm limit}&=& \left( \frac{{\rm SNR}\cdot 2k_B T_N }{\eta^2 f_c^{-1} R^2 d ~\rho_{\rm DM} B_0^2\sqrt{\Delta{t}}}\right)^{1/2} \left(\frac{m_a}{2\pi Q_c}\right)^{3/4}.
\eea
\cs{Here SNR$=3$ corresponds to a minimal 3$\sigma$ (99.6\%) credence level to differentiate the signal from noise. %Using SNR$=3$ is the common practice with Haloscopes (see e.g. ADMX~\cite{ADMX:2021nhd}, HAYSTAC~\cite{Zhong:2018rsr}).
}

\begin{figure}[ht]
\includegraphics[scale=0.15]{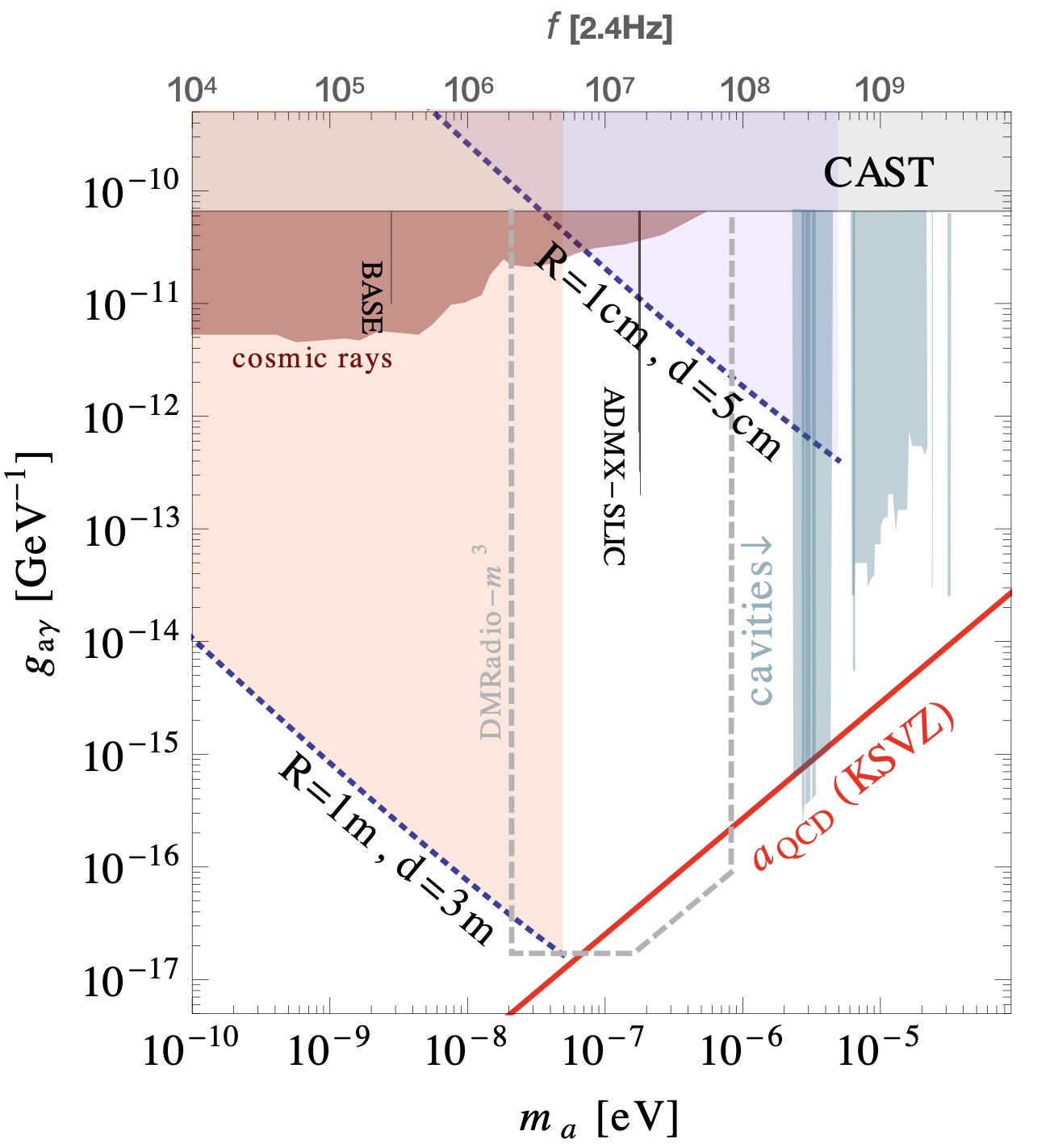}
\caption{Projected parameter space of sensitivity at benchmark setups (dotted curves), assuming $B_0=14$ T and one week's observation time at each frequency. The CAST exclusion limit \cite{Anastassopoulos:2017ftl} and recent constraints from resonance cavity experiments (stripes, right), cosmic rays (left, see eg. \cite{Kohri:2017ljt}), and non-cavity narrow-band experiments, such as BASE~\cite{BASE}, ADMX-SLIC~\cite{ADMXSLIC} and DMRadio-m$^3$(projected sensitivity) \cite{DMradio} are shown for comparison.
\cs{The frequency axis is transformed via $f = {m_a}/{2\pi}\simeq 
2.4$GHz$\big(m_a/10^{-5}$eV$\big)$.}}
\label{fig:reach}
\end{figure}

\cs{The $T_N=0.1$K estimate corresponds to a 30 dB cryogenic amplifier with 40K noise temperature, and it is a realistic benchmark within mature technology~\cite{4359086}. Lowering $T_N$ requires more powerful amplification and/or reducing amplifier input noise. Interestingly, recent developments with dual-path interferometry readout show the potential of significantly reducing the contamination from thermal noise~\cite{Bozyigit2010,Peng2016,Yang:2022uil}, thus lower $T_N$ can be possible. The temperature $T_c$ at the LCR circuit should be maintained below $T_{\rm AMP}/G$, which is in the order of 10 mK and within the capability of modern dilution fridges.}

We will consider two benchmark setups: (a) a modest-size setup of $R=1$ cm plates separated at $d=5$ cm as one scenario within the dimensions of a mainstream dilute fridge; (b) a more aggressive scenario with $R=1$ m, $d=3$ m to estimate the reach of this experimental method. We assume a magnetic field $B= 14$ T, $T_N=0.1$ K, and one week of measurement time at each frequency point in both benchmarks. The projected sensitivity limits on $g_{a\gamma}$ are shown in Fig.~\ref{fig:reach}.

Due to the $(\omega R)^2$ behavior in $\eta$, the expected $g_{a\gamma}$ limit weakens steeply towards lower frequencies, as shown in Fig.~\ref{fig:reach}. Thus the best limit is obtained for an $m_a$ near the coherence cut-off frequency, which decreases with larger experimental sizes. At benchmark (a), with a multi-centimeter sized magnetized region and $T_N$ at 1 K, the projected sensitivity is $g_{a \gamma}\sim 10^{-12}$ GeV at $m_a=5 ~{\rm \mu eV}$. For benchmark (b), reaching the QCD axion parameter space (red line) requires a multi-meter volume, which also leads to a lower cut-off $m_a$ around $50$ neV.

For a magnetized region of several centimeters, the maximal $m_a$ falls in the optimal detection range of cavity haloscopes. For a given setup dimension, $m_a$ below the cut-off frequency can also be measured at the cost of a smaller $\eta(\omega)$, and the electric probe can still provide a sensitive region of parameter space lower than the current solar axion limit~\cite{Anastassopoulos:2017ftl} at frequencies lower than cavity experiments. At $m_a$ above the cut-off, the sensitivity is expected to drop due to partial cancellation as the induced field can change the sign in the magnetized region. In addition, in the case of very low frequency $\omega R\ll 1$ where $\eta(\omega)$ suppression becomes severe, one may also consider different geometries of the electric plates, e.g. co-axial cylindrical plates to measure the radial component $E_r$ of the signal electric field, which can be of interest for future research. Our axion detector here is also ready to be a high-frequency gravitational wave detector, after some data re-analysis, which is a future direction \cite{Sun:2020gem,Aggarwal:2020olq,Berlin:2021txa,Domcke:2022rgu}.

Finally, we note the effectiveness of pursuing the electric signal for axion detection. As the time-varying effective $\vec{j}_a$ induces both electric and magnetic signals, both are viable detection options and the signals strongly depend on the geometry layout of the particular design. For a conceptual comparison of different methods, the key point of effectiveness is whether a high signal power $P_{\rm sig}$ can be realized for given apparatus dimensions and detector noise level. From Eq.~\eqref{eq:sig_power}, our maximal axion-conversion power is achieved at the cut-off wavelength $(\pi r^2 d)\sim (\pi/m_a)^3$, or $P_{\rm max}={\cal O}(1)\cdot Q_c\cdot\pi^3g_{a\gamma}^2\rho_{\rm DM} B_0^2/m_a^4$, which is comparable to the coherent axion-photon conversion power inside an effective magnetized volume. This $P_{\rm max}$ upper limit is comparable to that in a magnetic-signal-based LC design~\cite{Sikivie:2013laa}, like in ADMX-SLIC~\cite{ADMXSLIC}. %In our setup, we induced $E$ signal 
\cs{In our proposed detection system, the axion induces an electric signal}
inside the magnetized region, this however leads to steeper frequency dependence in the form factor $\eta(\omega)$ towards low frequency. We give a detailed comparison of signal powers in the Appendix.%~\ref{app:E-B}.

\bigskip
\bigskip
%\vspace{-10pt}
\begin{acknowledgments} \vspace{-10pt}
The authors thank for the support from the National Natural Science Foundation of China (Nos. 12150010, 12105013 and 12005255) and International Partnership Program of Chinese Academy of Sciences for Grand Challenges (112311KYSB20210012).
%Y.G. is supported under grant No. 12150010 supported by the National Natural Science Foundation of China. \ckk{Your grants}

%\bigskip
%\bigskip
\iffalse
{\bf{Conflict of interest:}} The authors declare that they have no conflict of interest.
{\bf{Author contribution:}} Junxi Duan and Yu Gao work out the experimental details. Chang-Yin Ji performs the numerical simulation. Yu Gao, Sichun Sun, Yugui Yao and Yun-Long Zhang initiate the project. All authors participate in the draft editing.
\fi
%\begin{figure}[h]
%    %\centering
%    \includegraphics[scale=0.4]{JiCY1.jpg}
%    \caption{    } \label{fig1}
%\end{figure}
\end{acknowledgments}

\appendix
%\section{Appendix}%\vspace{-10pt}

%\iffalse
%\section{In the  standard international(SI) unit}
%Refer to the conversion in \cite{Tobar:2018arx}, 
%The Lagrangian\begin{align}
%{\cal L} = -\frac{1}{4} F_{\mu\nu}F^{\mu\nu}-\frac{g_{a\gamma}}{4} a F_{\mu\nu}\tilde{F}^{\mu\nu}.
%\end{align}
%\section{Standard international(SI) unit}
\bigskip
{\section{Standard international(SI) unit}} 
We introduce the electric flux density $\vec{D}={\e}\vec{E}+\vec{P}$ and magnetic field intensity $\vec{H}=\vec{B}/{\m}-\vec{M}$,
where $\vec{P}$ is the polarization and $\vec{M}$ is the magnetization. The light velocity $c=1/\sqrt{\e\m}$. 
The modified Maxwell's equations in SI units are
\begin{align}
\vec\nabla\cdot \vec D&=
\rho_e-g_{a\gamma}\sqrt{\frac{\e}{\m}}\vec B\cdot \vec\nabla a, \nonumber\\
\vec\nabla\times\vec H-{\partial \vec D \over \partial t}
&=\vec j_e - g_{a\gamma}\sqrt{\frac{\e}{\m}} \left( \vec E\times \vec \nabla a-\vec B{\partial a\over  \partial t} \right).
%\vec\nabla\cdot\vec B&=0\nonumber\\
%\vec\nabla\times \vec E&=-{\partial \vec B\over \partial t}~,
\end{align}
%where  $\ga\equiv g_{a\gamma} \sqrt{\frac{\e}{\m}}$. 

%= g_{a\gamma} \e c
% For our case $\rho_e=0$, $j_e=0$, $\vec{E}=0$, then:
%\begin{align} \vec\nabla\cdot \vec D = -\ga\vec B\cdot\vec\nabla a &\equiv \rho_a, \nonumber\\
%\vec\nabla\times\vec H-{\partial \vec D \over \partial t} = \ga\vec B\, {\partial a\over  \partial t} &\equiv \vec{j}_a. \end{align}
%\fi

%{\bf{EM Simulation:}}
%\label{sect:sim}
%\section{Electromagnetic simulation}
%{\bf{Electromagnetic simulation}} ---
%\label{sect:sim_setup}
We compare our numerical in 2D with the analytical result, with different solenoid sizes and axion wavelengths. The boundary condition is set to be the perfect matching condition, and the simulation is done in the frequency domain in COMSOL. The 3D simulations are done with $10^4$ nodes.

\bigskip
%\vspace{10pt}
%\section{Capacitance on plates}
{\section{Capacitance on plates}} 
%\label{app:cap_plates}
Note that $E_z(r)$ varies in magnitude with $r$ and in phase. This is a complication at large frequencies that plates do not act purely as a capacitor. The phase difference between the plate's charge $q=\int E_z{\rm d}A$ and the voltage at the connection point $\Delta V=\left.E_z\right|_{r=0}\cdot d$ is
\bea 
\Delta\phi &=& {\rm Arg}\left[ i\pi J_1(\omega R)H_1^+(\omega R) - 1 \right] \nonumber \\
 &&- {\rm Arg}\left[ \alpha(\omega)J_0(0)-1\right]. 
\eea

\begin{figure}[h]
\includegraphics[scale=0.43]{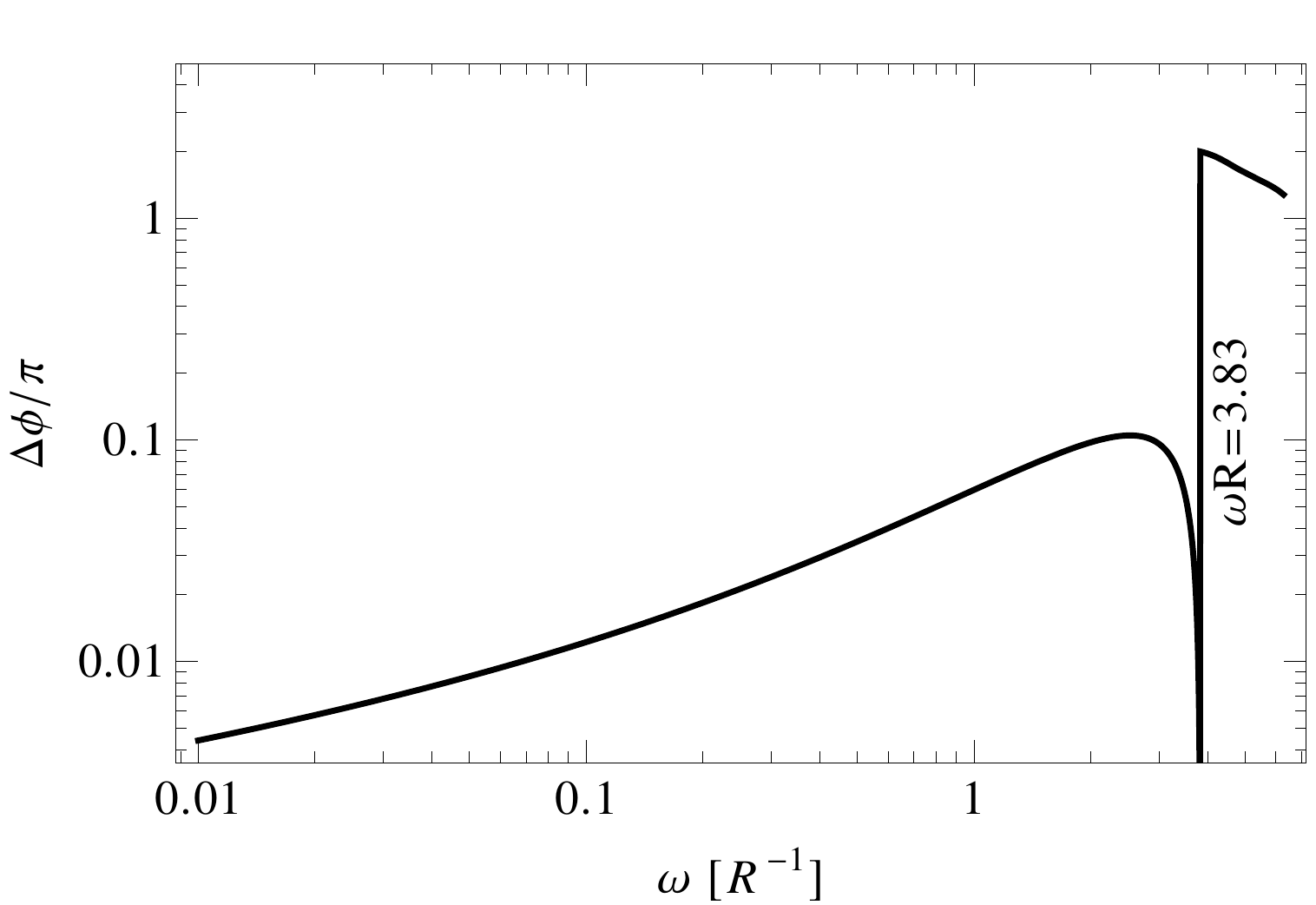}
\caption{Phase difference mod$(\Delta \phi/\pi,2)$ between the charge and voltage difference on the parallel plates, in terms of $\omega R$.}
\label{fig:phase_diff}
\end{figure}

As shown in Fig.~\ref{fig:phase_diff}, $\Delta\phi$ is significant near $\omega R \sim 1$, yet it diminishes towards $\omega R\ll 1$, so that the plates behave like a pure capacitor at the low frequency. When $\Delta\Phi$ is small, the plates' effective capacitance $C_1= q_0 /U_{r=0}$ can be estimated as
\be 
C_1= \frac{\pi R^2}{d}\cdot \frac{\vev{E}}{E_{r=0}}\equiv f_c(\omega)\cdot\frac{\pi R^2}{d},
\ee
and by evaluating from the expression for $E_{a,z}$, $f_c\sim 0.9$ at $\omega=R^{-1}$ and $f_c$ smoothly approaches to unity towards lower $\omega$, and the vanilla $\pi R^2/d$ is still a reasonable capacitance estimate at low frequency. For $r=1$ cm and $d=5$ cm, $C_1$ is $10^{-3}$ pF, and for a meter-scale setup $2R\sim d=1$ m, $C_1 =0.2$ pF. We will assume the plates dominate the total capacitance in the LC circuit, $C\approx C_1$, and a small $\Delta \phi$ can be compensated for by tuning the LC circuit to its resonance point.  With a quality factor $Q_c= 10^{-6}$ matched to that of the axion field, the resonant resistance $R_s=1/(Q_c\omega C)\sim 0.2~\Omega$ and $0.002 ~\Omega$ for $m_a =10^{-5}$ eV. 
%

%\vspace{10pt}
\bigskip
{\section{Comparison with a magnetic signal}} 
%\label{app:E-B}
Experimental sensitivities strongly depend on many factors, like the size of the magnetized volume, actual detector sensitivities, etc. Here we discuss our signal power and the magnetic signal design power~\cite{Sikivie:2013laa} as a conceptual comparison of the signal pickups with the electric/magnetic signal. Our signal power in Eq.~\eqref{eq:sig_power} can be written in resemblance to that in a cavity haloscope,
\bea
P_{\rm sig} = {\cal C} Q_c\cdot  g_{a\gamma}^2 \cdot \frac{\rho_{\rm DM}}{m_a}  B_0^2  \cdot V, 
\eea
where ${\cal C}=\eta^2f_c^{-1}$ denotes the geometric factor and $V=\pi r^2 d$ is the volume between the plates. This agrees with the total resonant coherent axion-photon conversion power inside a magnetized volume $V$ up to a factor ${\cal C}$. At the half-wavelength cut-off, e.g. $V\sim (\lambda/2)^3=(\pi/m_a)^3$, our ${\cal C}_{\rm max}$ evaluates to around unity at $\omega R=\pi/2$, and this leads to the maximal signal power
\be 
P_{\rm sig} < {\cal O}(1)\cdot Q\cdot\pi^3\cdot \frac{g_{a\gamma}^2\rho_{\rm DM} B_0^2}{m_a^4},
\ee
and it decreases when $\omega$ decreases, following the long-solenoid solution in Eq.~\eqref{eq:Ez}.

For a simplified comparison, LC~\cite{Sikivie:2013laa} and the later ADMX-SLIC~\cite{ADMXSLIC} design use a 2D loop as the magnetic signal pickup. If we ignore detector details and focus on the signal power in the pickup system, the LC pickup loop encloses a magnetic flux $\Phi_a=(g_{a\gamma}\sqrt{2\rho_{\rm DM}})B_0\cdot \frac{1}{4}l_m r_m^2$ and the resonance signal power is 
\bea
P_{\rm sig}^{\rm LC}&=&Q \frac{\vev{\Phi_a^2}}{L_m} \omega \nonumber  \\
&\approx& Q\cdot \frac{g_{a\gamma}^2\rho_{\rm DM}B_0^2}{l_m\ln{\left(\frac{r_m}{a_m}\right)}/\pi}\cdot m_a \cdot \left( \frac{1}{4} l_m r_m^2\right)^2,
\eea
where $l_m, r_m$ denotes the dimensions of the loops and $a_m$ is the loop's wire radius, following the notations in Ref.\cite{Sikivie:2013laa}. As it is a 2D pickup loop, this formula is without a volume and a geometric factor, so it is not easy to compare it with a cavity haloscope. This 2D loop is also limited by the cut-off wavelength. For $l_m, r_m \le \lambda/2$, we have 
\be 
P_{\rm sig}^{\rm LC} < Q\cdot \frac{\pi^6}{16\ln{\left(\frac{r_m}{a_m}\right)}}\frac{g_{a\gamma}^2\rho_{\rm DM}B_0^2}{m_a^4},
\ee
which also approaches the coherent axion-photon conversion power cap inside a cut-off sized volume $V=(\pi/m_a)^3$, up to a factor of the same order of magnitude.

This comparison shows that, at optimal frequencies, both $E$ and $B$ signal-based methods can reach up close to the coherent axion-photon conversion power with a matched effective volume. For frequency dependence, our geometric factor has a stronger dependence on frequency and the best sensitivity is reached at the largest allowed $m_a$.

\bibliography{refs}

\end{document}